\documentclass[journal]{IEEEtran}
\usepackage{cite}
\usepackage[pdftex]{graphicx}
\usepackage{amsmath}
\usepackage{lipsum}
\usepackage{color}
\usepackage{soul}
\usepackage{balance}

\begin{document}

\title{	Low Barrier Magnet Design for \\ Efficient Hardware Binary Stochastic Neurons 
\author{Orchi Hassan$^1$, Rafatul Faria$^1$, Kerem Y. Camsari$^1$, Jonathan Z. Sun$^2$ and Supriyo Datta$^1$} 
\thanks{OH, RF, KYC, SD are with the School of Electrical and Computer Engineering, Purdue University, West Lafayette, IN, 47906 USA. JZS is with IBM T. J. Watson Research Center, Yorktown Heights, NY 10598 USA}
\thanks{Manuscript received XX, 201X; revised XX, 201X.}}
\maketitle

\begin{abstract}
Binary stochastic neurons (BSN's) form an integral part of many machine learning algorithms, motivating the development of hardware accelerators for this complex function. It has been recognized that hardware BSN's can be implemented using low barrier magnets (LBM's) by minimally modifying present-day magnetoresistive random access memory (MRAM) devices. A crucial parameter that determines the response of these LBM based BSN designs is the \emph{correlation time} of magnetization, $\tau_c$. In this letter, we show that for magnets with low energy barriers ($\Delta \approx k_BT$ and below), circular disk magnets with in-plane magnetic anisotropy (IMA) lead to $\tau_c$ values that are two orders of magnitude smaller compared to $\tau_c$ for magnets having perpendicular magnetic anisotropy (PMA) and provide analytical descriptions. We show that this striking difference in $\tau_c$ is due to a precession-like fluctuation mechanism that is enabled by the large demagnetization field in IMA magnets. We provide a detailed energy-delay performance evaluation of previously proposed BSN designs based on Spin-Orbit-Torque (SOT) MRAM and Spin-Transfer-Torque (STT) MRAM employing low barrier circular IMA magnets by SPICE simulations. The designs exhibit sub-ns response times leading to energy requirements of $\sim$a few fJ to evaluate the BSN function, orders of magnitude lower than digital CMOS implementations with a much larger footprint. While modern MRAM technology is based on PMA magnets, results in this paper suggest that low barrier circular IMA magnets may be more suitable for this application. 
\end{abstract}


\IEEEpeerreviewmaketitle
\vspace{-15pt}
\section{Introduction}
\vspace{-5pt}
Many inference and machine learning algorithms are based on networks of binary stochastic neurons (BSN's)\cite{ackley1985learning,amit1992modeling,BSNr2rt,alaghi2013survey, esser2013cognitive, merolla2014million} each of whose response  $m_i$ at time step (n+1) is determined by  the input $I_i$ at time n ($r_i$: random number between $-$1 and $+$1):
\vspace{-2.5pt}
\begin{equation} \label{BSN:eq1}
m_i (n+1) = {\rm{sgn}}[ \mathrm{tanh} \ {I_i(n)} - r_i]
\end{equation}
In the absence of an input $I_i$ the output $m_i$ fluctuates randomly between two values $-$1 and $+$1. A positive $I_i(n)$ makes +1 more likely, while a negative $I_i(n)$ makes $-$1 more likely \cite{footnote1}.  Each BSN described by Eq.~\ref{BSN:eq1} receives its input from a weighted sum of other BSN's obtained from a ``synapse'' $   {I_i(n)} = \sum_j{W_{ij} \ m_j(n)} $. A wide variety of functions can be implemented by properly designing or learning the weights $W_{ij}$ \cite{jo2010nanoscale,ccilingiroglu1991purely, hassan2018voltage}.

The BSN function (Eq.~\ref{BSN:eq1}) is evaluated repeatedly in modern algorithms but they are typically implemented in software. Efforts have been put into developing a suitable hardware for accelerating evaluation of this function, many of which are based on magnetoresistive random access memory (MRAM) technology which is a major contender in the field of non-volatile memory using stable magnets to store information in the form of $0$'s and $1$'s. By contrast, BSN's can be built out of nanomagnets designed to have low energy barriers \cite{zink2018telegraphic, parks2018superparamagnetic, vodenicarevic2017low, mizrahi2018neural, vodenicarevic2018circuit, sutton2017intrinsic, faria2017low, liyanagedera2017stochastic}. The performance of such BSN designs are largely dependent on the magnetization fluctuation rates of the LBM's, making it important to design the low barrier magnet to have a high fluctuation rate. 

Stable magnets could be redesigned to have low energy barriers  by scaling the magnetic anisotropy \cite{debashis2018design}. The energy associated with a magnet is given by
\vspace{-2pt}
\begin{equation*}
 E =\frac{1}{2} H_{kp}M_s\Omega (1-m_x^2)+\frac{1}{2}H_{ki}M_s\Omega (1-m_z^2)
\end{equation*}
where,  $H_{kp}=2K_s/t-4\pi M_s$ is the perpendicular anisotropy field along x-axis, $K_s$ is the surface anisotrpy density, $ H_{ki}$ is the in-plane anisotropy along z-axis, $M_s$ is the saturation magnetization and $\Omega$ is the volume of the magnet.  Low barrier magnets can be obtained by adjusting the thickness $t$ of perpendicular anisotropy (PMA) magnets so that $H_{kp}\approx0$ making $\Delta_{PMA}=H_{kp}M_s\Omega /2 \approx 0$ or by making in-plane anisotropy (IMA) magnet's shape circular so that $H_{ki}\approx 0$ making $\Delta_{IMA}=H_{ki}M_s\Omega/2 \approx 0$. Such magnets with diameters that are less than about 100 nm have been  shown to exhibit monodomain behavior \cite{debashis2018design, cowburn1999single, debashis2016experimental}. It is important to note that while modifying existing interfacial PMA free layers by modulating the thickness to make them IMA seems relatively straightforward, replacing highly optimized fixed PMA layers \cite{park2015systematic} with IMA stacks could prove more challenging. 

\begin{figure}[!tb]
\centering
\includegraphics[width=0.99\linewidth]{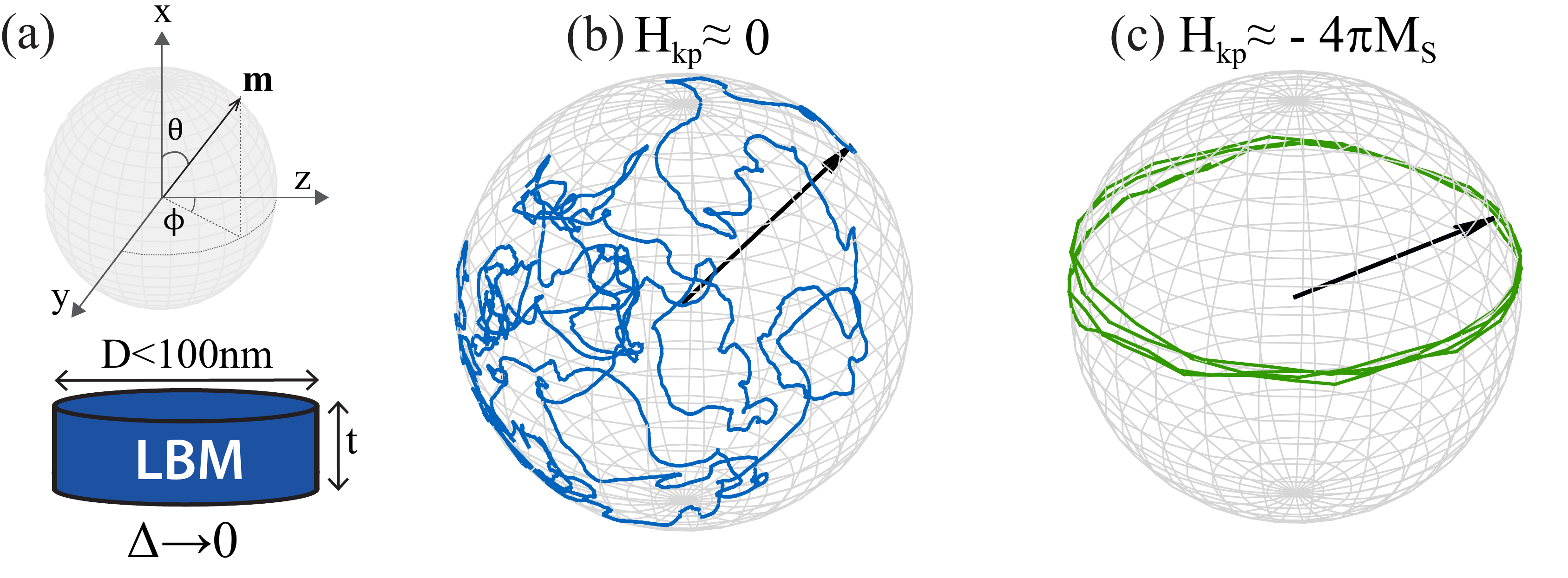}
 \caption{\textbf{Fluctuation Dynamics of LBM:} (a) Schematic illustration of circular LBM with saturation magnetization $ M_s$ and volume $ \Omega=\pi(D/2)^2t$ and the magnetization $\textbf{m}=\textbf{M}/M_s=(m_x,m_y,m_z)\equiv (\cos\theta, \sin\theta \sin\phi, \sin\theta \cos\phi)$. SPICE simulation shows $m(t)$ dynamics on Bloch sphere of a low barrier circular magnet with ($\Delta \approx 0$)  for magnet with (b) $ H_{kp} \approx 0$ and (c) $\rm H_{kp}\approx-4\pi M_s\approx-13.8~kOe$, where $ H_{kp}=2K_s/t-4\pi M_s$ is the perpendicular anisotropy along x-axis and the in-plane anisotropy $ H_{ki} \approx 0$ due to circular shape. \vspace{-15pt}}
\label{fi:figLBMm}
\end{figure}

The time scale of fluctuations can be very different for the two categories of low barrier magnets as shown in Fig.~\ref{fi:figLBMm}b and c. In PMA with vanishing perpendicular anisotropy field making $\Delta \rightarrow 0$, the thermal noise makes the magnetization fluctuate randomly anywhere on the Bloch sphere, while in circular IMA with no preferred easy axis and a large effective demagnetization field ($H_D=4\pi M_s$) restricts the fluctuations to  to a compressed region near the equator (i.e. in-plane moment), making more rapid fluctuations possible.

In this letter, we present a distinction between fluctuation dynamics of low barrier PMA and IMA magnets providing analytical expressions for two very important parameters for performance evaluation of hardware BSN's: the correlation time $\tau_c$ and pinning current $I_p$ for $\Delta \approx k_BT$ and below. Circular IMA magnets have a correlation time two orders of magnitude smaller compared to PMA and a pinning current that is much higher. We also present a device level performance evaluation on two previously proposed compact BSN designs \cite{camsari2017stochastic, camsari2017implementing} using circular IMA magnet and show that the sub-ns operation results in only $\sim$ a few fJ of energy requirement for evaluating the BSN function which is orders of magnitude lower than its CMOS implementation \cite{ardakani2017vlsi,yuan2017vlsi}.

\vspace{-5pt}
\section{Low barrier magnets}

Binary stochastic neurons could be viewed as a tunable random number generator and a key parameter defining its performance would be the rate at which it produces the random numbers. For an LBM BSN, this rate is related to the magnetization fluctuation rate of the low barrier magnet. The time it takes for the magnet to lose its memory, the \textit{correlation time} $\tau_c$ is defined by the full-width-half-maxima of the temporal auto-correlation function $C(t)$ of magnetization and could be used to characterize the relevant time-scale of operation of BSN.

\begin{figure}[!b]
\centering
\includegraphics[width=0.99\linewidth]{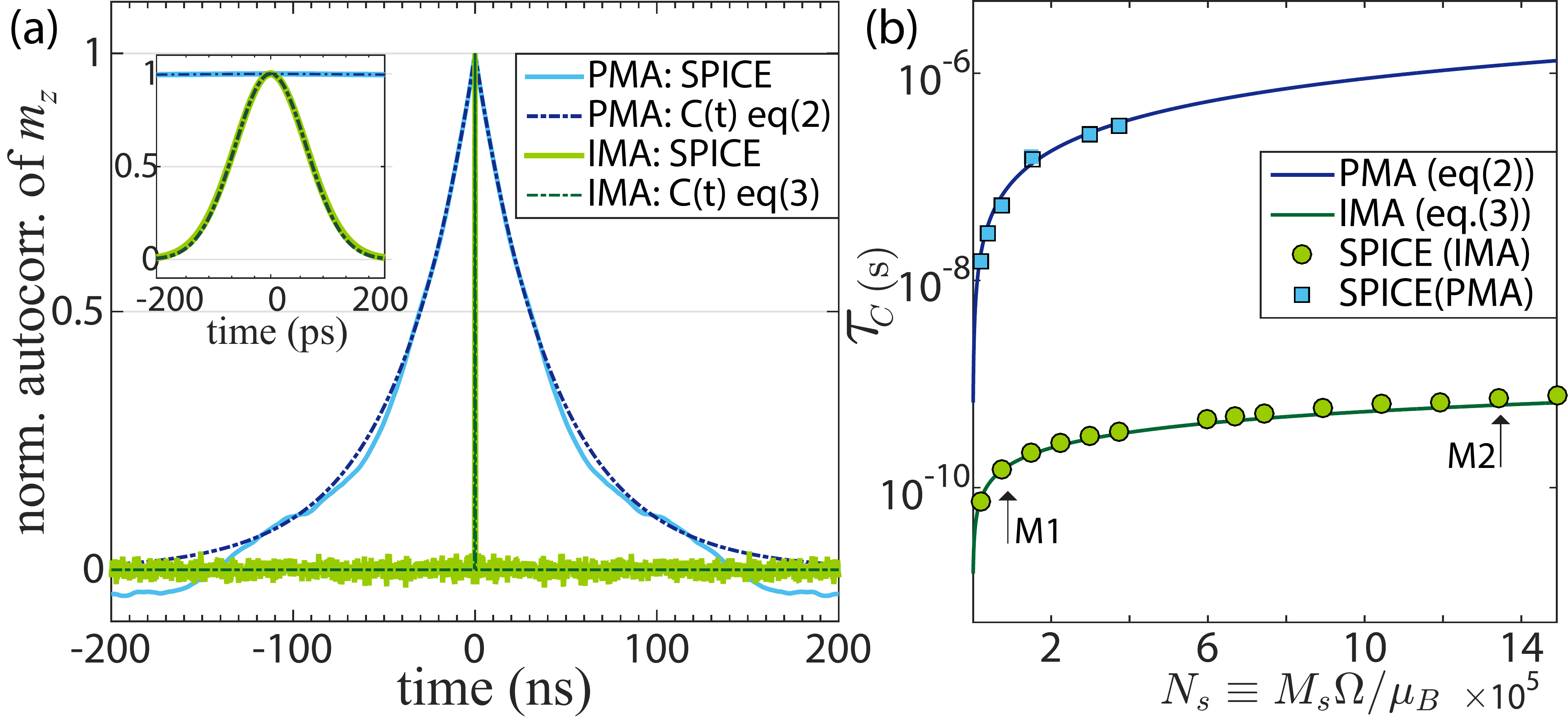}
 \caption{\textbf{Correlation Time of PMA and IMA magnets} (a) The normalized auto-correlation of magnetic fluctuations taken in the z direction, (b) Comparison of $\tau_c$ as a function of number of spins $N_{s} \equiv M_{s}\Omega/\mu_{B}$ where $M_s=1100 \ \rm emu/cc$ and the volume $\Omega$ is varied. Damping coefficient $\alpha$ is assumed to be 0.01: Results from numerical simulations agree well with the equations cited in the text. \vspace{-20pt}}
\label{fi:figLBMtauC}
\end{figure}
In low barrier magnets where the energy barrier is well below the thermal energy ($\Delta \ll k_BT$) its magnetization becomes a continuous variable. The Arrhenius law which describes the thermal fluctuations of high barrier magnets ($\Delta \gg  k_BT$) with two distinct magnetic states thus does not hold for LBM \cite{faria2017low, lopez2002transition}. Instead, thermal fluctuations in monodomain low barrier magnets could be characterized starting from Fokker-Planck equation (FPE)\cite{brown1963thermal, coffey2012thermal} or the Landau-Lifshitz-Gilbert (LLG) equation including a Langevin term describing thermal fluctuation \cite{lopez2002transition, kaiser2018ultrafast}.

Coffey et. al. \cite{coffey2012thermal} analyzes the magnetic fluctuations in a PMA magnet due to thermal noise in detail by using the Fokker-Planck equation (FPE) derived by W. F. Brown \cite{brown1963thermal}. The analysis presented in these references focused on high-barrier magnets but are not limited to it and thus can be evaluated for $\Delta \rightarrow 0$ to describe the low barrier magnet dynamics of PMA magnets which agree well with numerical results. 
\vspace{-10pt}
\begin{equation} \label{TD:PMA}
\begin{aligned}
\text{PMA:} \ \ C(t) & = \mathrm{exp} \left(-2\alpha \gamma \frac{k_BT}{M_S\Omega} |t|\right) \\
	\tau_{c} & = \frac{M_S \Omega}{\alpha \gamma k_BT} \mathrm{ln}(2) 
\end{aligned}
\end{equation}
In low barrier circular IMA magnets when thermal noise kicks the magnetization out-of-plane, due to absence of an easy axis and the presence of large orthogonal demagnetization field $H_D$ the in-plane magnetization starts precessing. If we consider an ensemble of such magnets each with a different precession frequency due to thermal noise, the average magnetization vector would quickly dissipate. The auto-correlation function of the in-plane magnetization $m_z=\cos(\phi(t))$ could be expressed as:
\vspace{-5pt}
\begin{equation*}
C(t)  = \displaystyle\int_{-1}^{1}dm_x \cos(\gamma H_D m_x t) \rho(m_x)\bigg/ \displaystyle\int_{-1}^{1}dm_x \rho(m_x)
\end{equation*}
where the in-plane precession dynamics is described by $\phi(t)\approx \gamma H_D m_x t$ \cite{kaiser2018ultrafast} for low damping $\alpha$. The perpendicular magnetization $m_x$ follows a Boltzmann distribution with $\rho(m_x)\approx \exp(-H_DM_S\Omega m_x^2/2 k_BT)$. For large values of $H_D$ the integral could be extended to $\pm \infty$ and evaluated to give an expression for the auto-correlation function and correlation time as follows:
\vspace{-5pt}
\begin{equation} \label{TD:IMA}
\begin{aligned}
	\text{IMA:} \ \ C(t) &= \mathrm{exp} \left( - \gamma^2 \left(\frac{H_D k_B T}{M_S \Omega} \right) \frac{t^2}{2} \right) \\
	 \tau_{c}&= \sqrt{8\ \mathrm{ln}(2)} \frac{1}{\gamma} \sqrt{\frac{M_S \Omega}{H_D k_BT}}
\end{aligned}
\end{equation}

\begin{figure}[!b]
\centering
\includegraphics[width=0.92\linewidth]{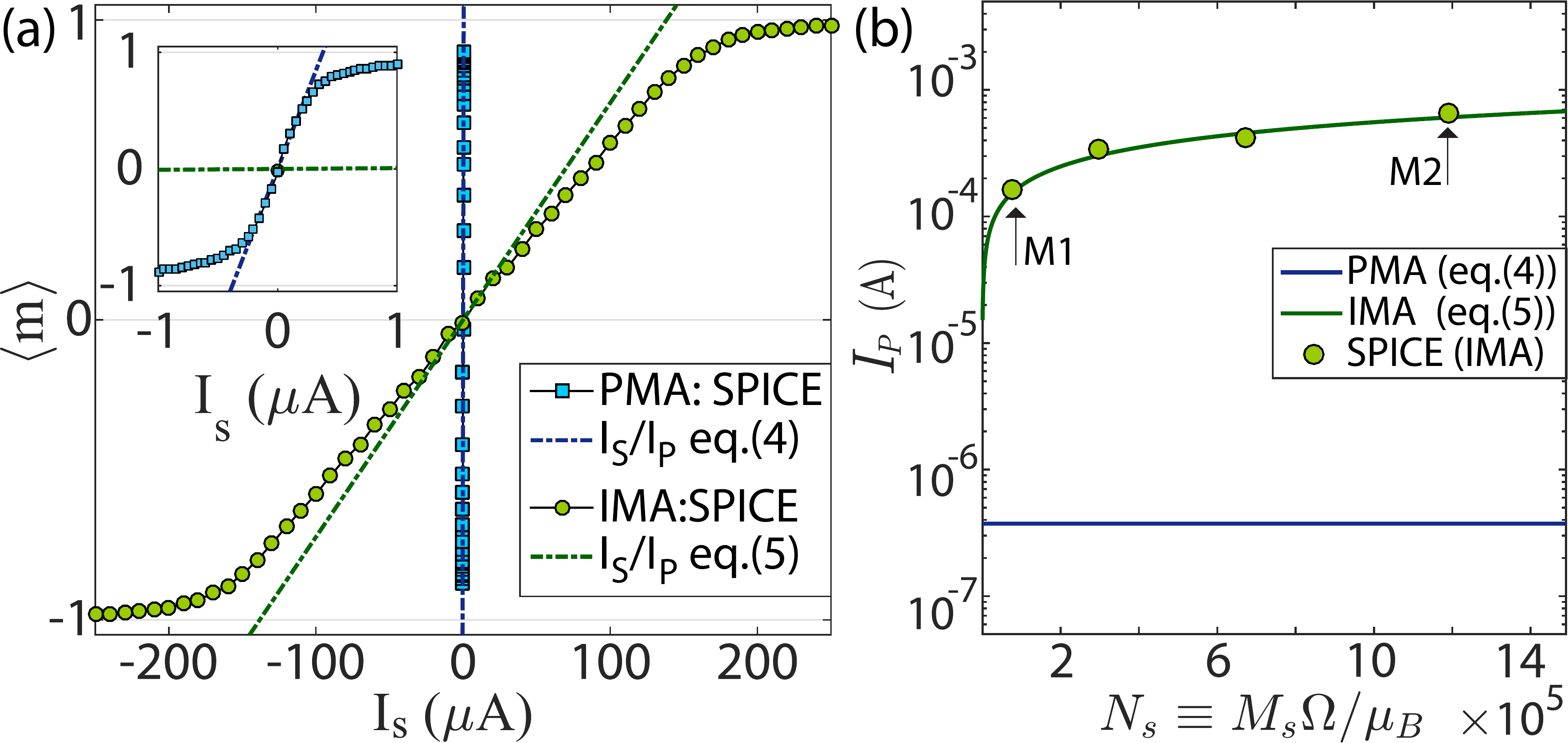}
 \caption{\textbf{Pinning current of PMA and IMA magnets} (a) PMA and IMA magnet's long time averaged magnetization $\langle m \rangle$ as a function of applied spin current $I_S$, (b) Comparison of PMA/IMA $I_P$ as a function of number of spins $N_{s} \equiv M_{s}\Omega/\mu_{B}$ where $M_s=1100 \ \rm emu/cc$ and the volume $\Omega$ is varied. Damping coefficient $\alpha$ is assumed to be 0.01: Results from numerical simulations agree well with the equations cited in the text. \vspace{-20pt}}
\label{fi:figLBMIp}
\end{figure}

In numerical simulations, we observe essentially the same auto-correlation behavior, even when the correlation function is obtained from the time-dependent fluctuations of a single magnet fluctuating for long time periods as shown in Fig.~\ref{fi:figLBMtauC}a. In PMA no such precessional fluctuation mechanism exists as the internal fields are compensated. 

Another important parameter for evaluating an LBM based stochastic device performance is itds sensitivity to spin current. To maintain stochasticity in MRAM type devices, they should be immune to read current, and the amount of current required to bias BSN devices is also relevant for power considerations. In high barrier magnets the concept of switching current is presented \cite{sun2000spin}, for low barrier magnets we refer to \textit{pinning currents} as the relevant quantity which can be mathematically defined as: $I_P = (\langle m \rangle / I_S)^{-1}$ as shown in Fig.~\ref{fi:figLBMIp}. The pinning currents for PMA can be derived from steady-state Fokker-Planck equation as described in Ref.~\cite{sayed2018zero}, while for IMA magnets with $ \Delta \rightarrow 0$ and low damping, the pinning current can be approximated from the relation $I_P \equiv {qN_S C(0)}\big/{\int_{0}^{\infty}dt C(t)}$. Fig.~\ref{fi:figLBMIp} shows that the numerical results are well described by the obtained expressions:
\vspace{-5pt}
\begin{align} \label{IP:PMA} 
\text{PMA:} \ \ \ I_{P}=  \frac{6q}{\hbar} \ \alpha k_BT \hspace{0.1in}
\end{align}
\vspace{-5pt}
\begin{equation} \label{IP:IMA}
\text{IMA:}  \\\ I_P= \frac{2q}{\hbar} \sqrt{\frac{2}{\pi}} \sqrt{H_D M_S \Omega \ k_BT} 
\end{equation}
The derivation of Eq.~\ref{IP:PMA} and Eq.~\ref{IP:IMA} assume zero energy barriers, but numerically we observe that these equations are approximately valid for barriers up to $\Delta \approx k_B T$. In practice obtaining near-zero barrier circular magnets could be challenging due to process variation. For interconnected networks of p-bits, a distribution of correlation times for each p-bit needs to be considered as shown in Ref.\cite{pervaiz2017hardware}.

Note that IMA-based designs can achieve sub-nanosecond correlation times even with fairly large volumes, provided that monodomain behavior can be preserved with a small enough diameter, while PMA-based designs tend to be much slower making IMA magnets more suitable for BSN applications. This is accompanied by fairly large pinning currents for IMA compared to PMA which minimizes read disturb effects.  

In the following section for the performance evaluation of two LBM based hardware BSN designs we used circular IMA magnets M1 and M2 with volumes  $ 800 \pi$ and $20480 \pi \ \rm nm^3$, respectively.

\vspace{-5pt}
\section{Performance Evaluation of Hardware BSN using circular IMA LBM}
\begin{figure}[!h]
\centering
\includegraphics[width=0.95\linewidth]{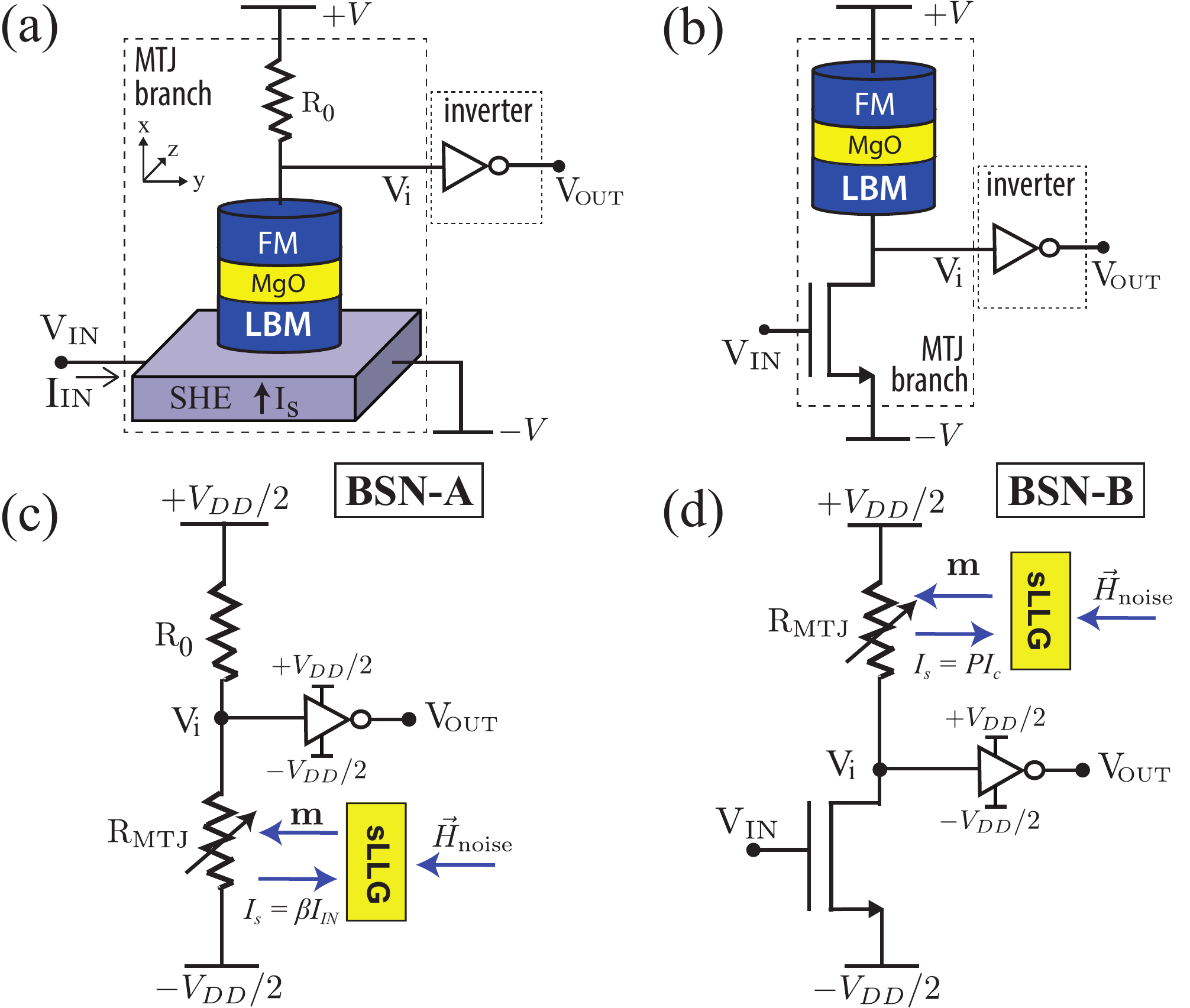}
 \caption{\textbf{Two BSN designs using stochastic MTJ with fluctuating resistance}: (a) \textbf{BSN-A} uses an input spin current to pin the fluctuating resistance \cite{camsari2017stochastic}. Structurally it looks similar to spin-orbit torque magnetoresistive random access memory (SOT-MRAM). (b) \textbf{BSN-B} looks similar to spin transfer torque MRAM (STT-MRAM) but it makes no use of spin torque. The input voltage controls the resistance of a field effect transistor (FET) which is in series with the MTJ \cite{camsari2017implementing}. (c) and (d) show the circuit models used for SPICE simulations. \vspace{-15pt}} 
\label{fi:fig1BSN}
\end{figure}

In this section we evaluate the steady-state and time response of two hardware BSN designs proposed in the past \cite{camsari2017stochastic, camsari2017implementing} shown in Fig.~\ref{fi:fig1BSN} and measure the energy and delay associated with each. 

The designs makes use of a magnetic tunnel junction (MTJ) whose free layer is a low barrier magnet with a fluctuating magnetization $m_{z}(t)$, resulting in a fluctuating resistance, $R_{\rm{MTJ}}(t)^{-1}=G_0[1+m_{zi}(t)\rm{TMR}/(2+\rm{TMR})]$ where $G_0$ is the average conductance and TMR is the tunneling magnetoresistance. The fluctuating resistance $R_{MTJ}(t)$ is converted to a fluctuating voltage $V_{i}$(t) by the potential divider:
\begin{equation}\label{eq6:BSNA_Voi}
\frac{V_{i}(t)}{V_{DD}/2}=(\pm) \frac{R_{\rm{MTJ}}(t)-R_0}{R_{\rm{MTJ}}(t)+R_0}
\end{equation} 
The fluctuations are controlled by two different mechanisms in the two designs. BSN-A is a spin-orbit-torque controlled device \cite{camsari2017stochastic} which uses the input spin current (in y direction) from the GSHE layer to pin the free layer magnetization (in z direction) of the MTJ thereby pinning $R_{MTJ}$ and implements ($+$) configuration of Eq.~\ref{eq6:BSNA_Voi}. BSN-B is a series resistance controlled device \cite{camsari2017implementing} which uses the input voltage to control the transistor resistance $R_{0}$ and implements the ($-$) configuration of Eq.~\ref{eq6:BSNA_Voi}. Ideally $R_{MTJ}$ remains unchanged, though in actual designs it may be important to consider unintended pinning effects of the current. Both designs use a minimum sized CMOS inverter to convert the fulctuating $V_i$ into a rail-to-rail output $V_{OUT}$. In each case we will use SPICE simulations based on state-of-the-art stochastic Landau-Lifshitz-Gilbert (s-LLG) models for LBM's \cite{camsari2015modular} free layer of the MTJ having $G_0 \simeq (25 K\Omega)^{-1}$ and $\mathrm{TMR}=2P^2/(1-P^2) = 110 \%$ with polarization $P\simeq0.6$ coupled with 14 nm HP FinFET's \cite{cao2002predictive} to show that the output voltage $V_{OUT}$  from a specific BSN is approximately related to its input $V_{IN}$ by an equation that mimics Eq.~\ref{BSN:eq1} :
\vspace{-2.5pt}
\begin{equation}\label{eq3}
   \frac{V_{OUT} (t+t_0)}{V_{OUT0}} \approx {\rm{sgn}} \bigg[ \mathrm{tanh} \ \frac{V_{IN} (t)}{V_{IN0}} - r (t) \bigg]
\end{equation} 
\noindent with scaling factors $V_{OUT0}, V_{IN0}, t_{0}$ characterizing the specific hardware design.

\vspace{-10pt}
\subsection{Steady-State Response}
\begin{figure}[!ht]
\centering
\includegraphics[width=0.99\linewidth]{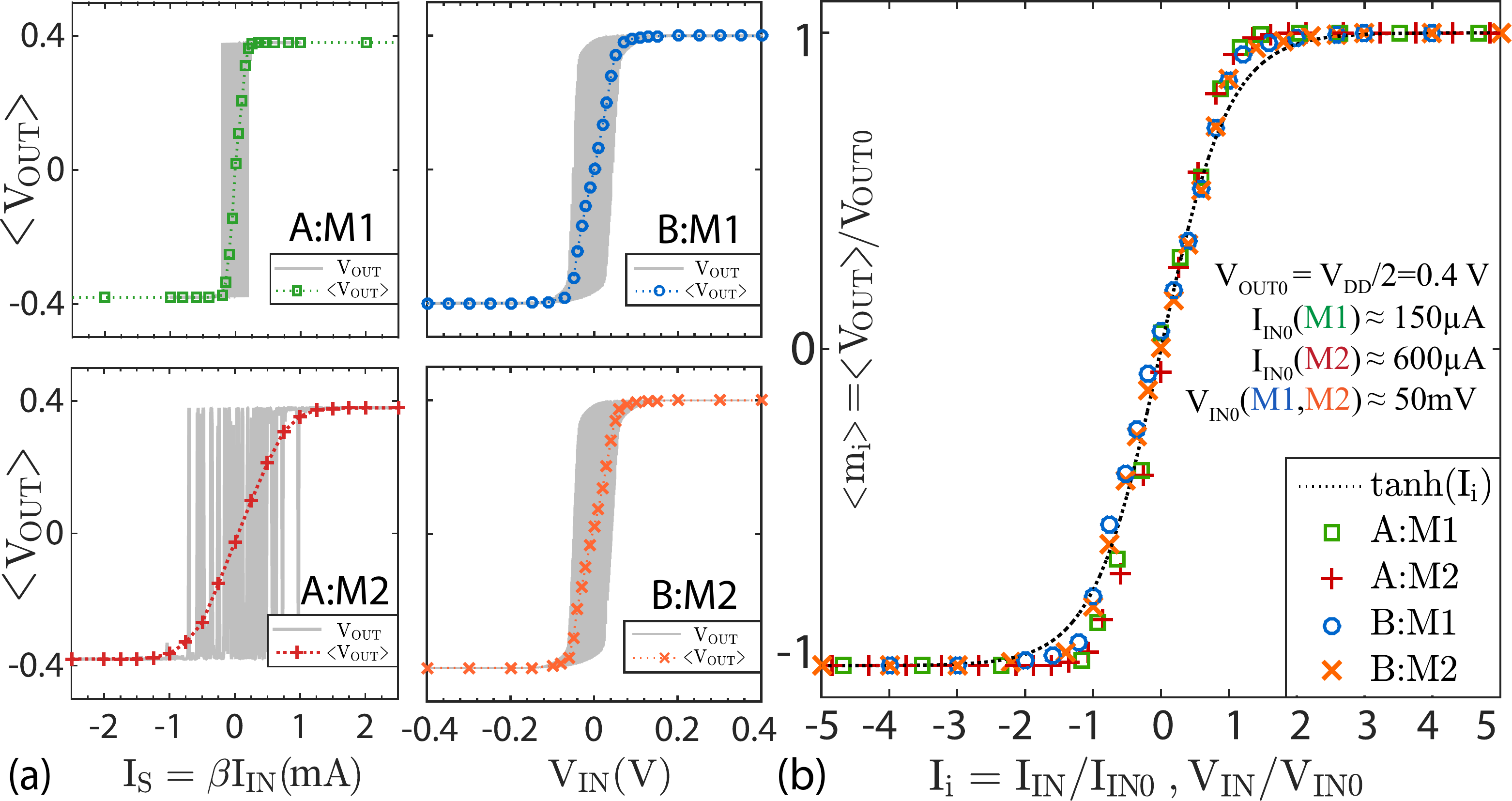}
 \caption{\textbf{Steady-state Response:} (a) Plot of $\rm{\langle V_{OUT}\rangle}$ (averaged over a time window $\gg \tau_c$) vs $V_{IN}$ for designs A, B using magnets M1, M2. The grey lines indicate $\rm{V_{OUT}}$ without time averaging.  (b) All four plots in (a) collapse onto a single curve using appropriate scaling parameters $\rm{V_{OUT0}, I_{IN0}, V_{IN0}}$. The resulting curve approximately follows the time averaged $\rm{\langle m_i \rangle }$ of Eq.~\ref{BSN:eq1}. \vspace{-17pt} }
\label{fi:fig3Sig}
\end{figure}

Fig.~\ref{fi:fig3Sig} shows the individual steady state response of design A,B using magnet M1 and M2, which can all collapse onto the same curve using appropriate scaling parameters. The output scaling quantity $\rm{V_{OUT0} \simeq V_{DD}/2 = 0.4V}$ is the same for all cases as this quantity is defined entirely by CMOS inverter output voltage swing. On the other hand, the input scaling parameters are very design dependent. For BSN-A $\rm{I_{IN0}}$ is determined by pinning currents of magnets M1 and M2. Indeed, the scaling parameters in Fig.~\ref{fi:fig3Sig}b were obtained from Eq.~\ref{IP:IMA}. For BSN-B $\rm{V_{IN0} \sim 50mV}$ for both magnets, determined by transistor characteristics. Note that the SPICE simulations include the read disturb current, but its effect is minimal due to the high pinning currents of low barrier IMA compared to PMA as can be seen from Eq.~\ref{IP:PMA} and Eq.~\ref{IP:IMA}. 
\vspace{-5pt}
\subsection{Time Response}
\begin{figure}[!h]
\centering
\includegraphics[width=0.95\linewidth]{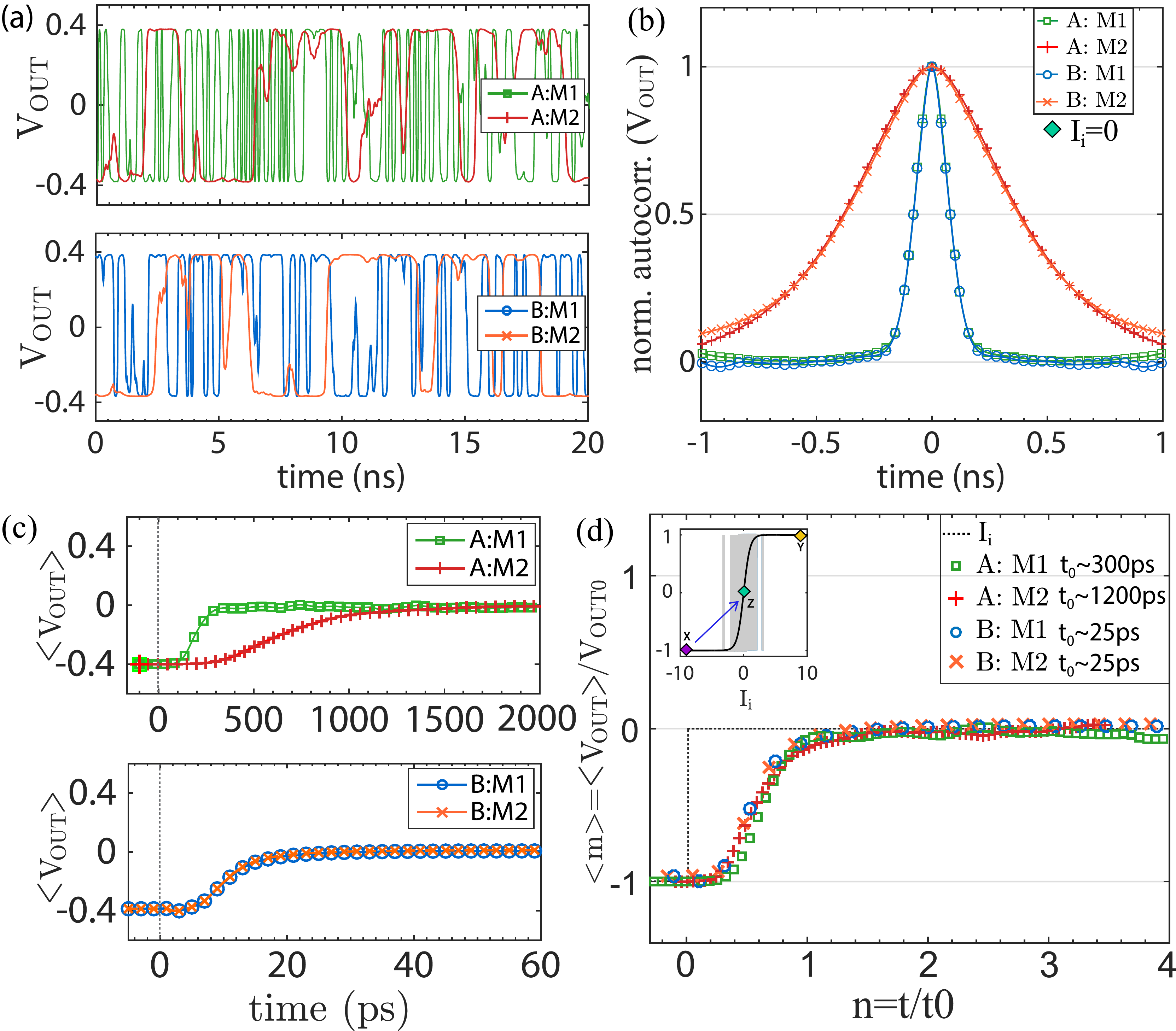}
 \caption{\textbf{Two relevant time-scales for BSN Operation:} (a), (b) show correlation time and (c),(d) show response time. (a) Output voltage fluctuations with $I_i=0$ for designs A, B using magnets M1, M2. (b) Corresponding normalized autocorrelation functions. (c) Response to a step function $I_i: -10 \rightarrow 0$ at t=0 averaged over 1000 ensembles for all four cases.(d) All four curves in (c) collapse onto a single curve using appropriate scaling parameter $t_0$.\vspace{-18pt}} 
\label{fi:fig4Time}
\end{figure}

Fig.~\ref{fi:fig4Time} shows the two relevant timescales associated with BSN operation. First is the correlation time of the output voltage which is determined by the magnet parameters. Indeed, the FWHM of the autocorrelation function corresponds well to Eq.~\ref{TD:IMA}, which is expected since circuit related times are much shorter in this case. Second is the response time which is very design dependent. For BSN-A it is determined by magnet physics while for BSN-B it is determined by transistor physics \cite{nikonov2015benchmarking}. Our analysis shows that the response time $\rm{t_0}$ of a single BSN-B neuron is independent of magnet parameters. However, the response of an interconnected network of such neurons would also involve the magnet correlation time $\tau_c$. 

\vspace{-5pt}
\subsection{Power Consumption}
\vspace{-5pt}
\begin{figure}[!ht]
\includegraphics[width=0.99\linewidth]{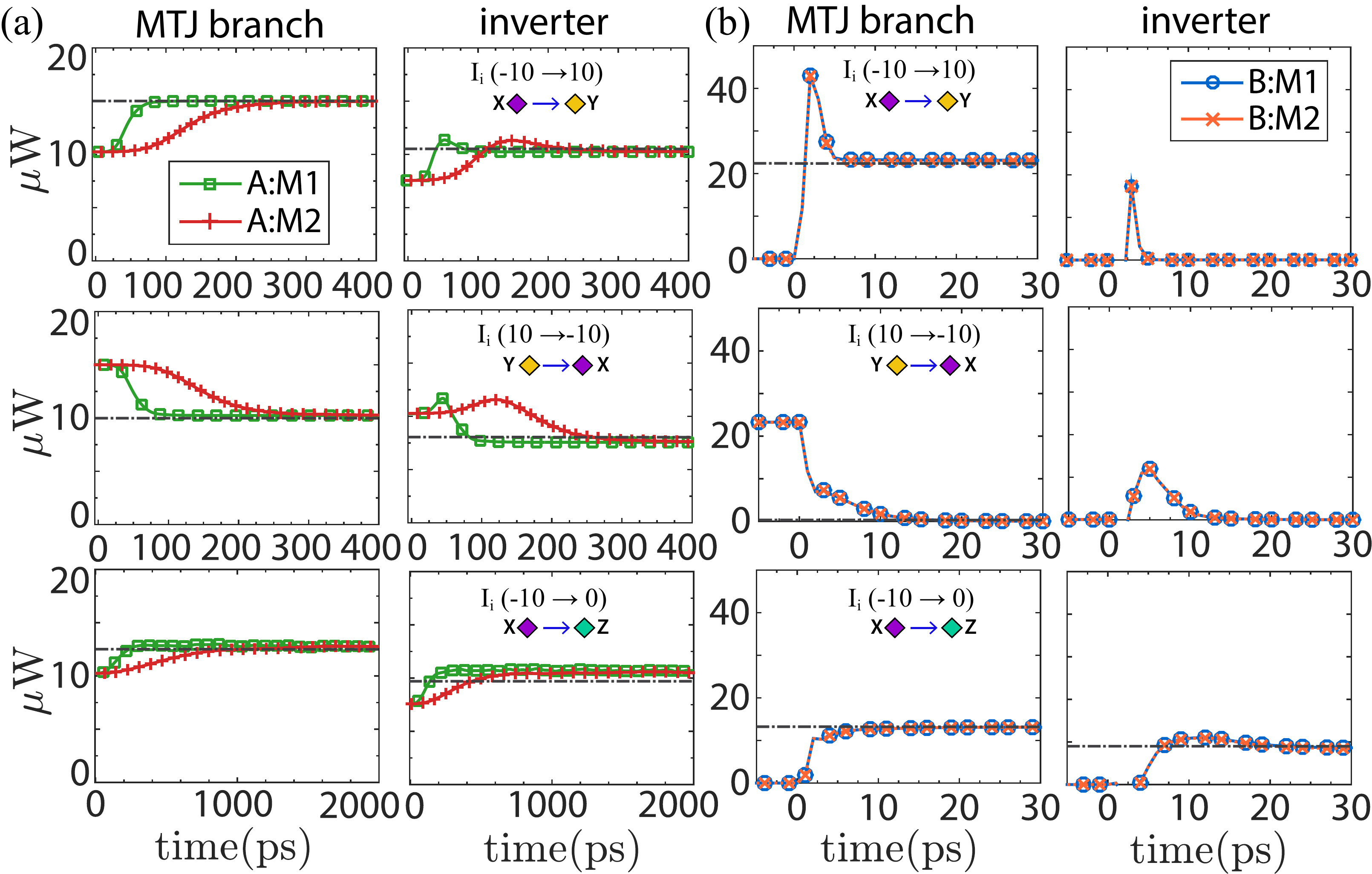}
\caption{\textbf{Power Consumption} for (a) BSN-A and (b) BSN-B when the input is stepped at t=0 as indicated.} \vspace{-20pt}
\label{fi:fig5Power}
\end{figure} 
Fig.~\ref{fi:fig5Power} shows the power drawn from the sources $\pm V_{DD}/2$ individually by the MTJ branch and the inverter branch as $V_{IN}$ is stepped at $t=0$ from different initial to final values as indicated. The steady-state values of the power dissipated in both the MTJ and inverter branches agree quantitatively with the simple estimate (see  dashed line in Figures)  from $V_{DD}^2/R$, where R is the appropriate resistance, namely $R_{MTJ} + R_0$ for the MTJ branch, and $R_{\rm NMOS}+R_{\rm PMOS}$ for the inverter branch. For the MTJ branch, the power dissipated is $\sim$10-20 $\mu W$ for all cases except in the middle panel for BSN-B. In this case the final state involves a large negative input voltage $V_{IN}$ for which the series transistor is turned OFF, making the resistance R extremely large, so that $V_{DD}^2/R \rightarrow 0$. In all other cases, the total R is of the order of the MTJ resistance $\sim 25 K\Omega$, so that $V_{DD}^2/R \sim 25 \mu W$. For the inverter branch, BSN-A dissipates $\sim$10 $\mu W$ since the voltage at the inverter input in all cases remains close to the threshold value making both NMOS and PMOS branches fairly conducting. On the other hand, for BSN-B, PMOS and NMOS get turned off for large positive and for large negative input $V_{IN}$ respectively, making the effective R very large. Only for input voltages $\sim 0$,  both PMOS and NMOS branches are conducting, giving rise to a steady-state power $\sim 10 \mu W$ like BSN-A. This number could be lowered if we can engineer larger voltage fluctuations at the inverter input, $ \lvert \delta V_i \rvert  \sim \ P^2 V_{DD}/(4-P^4)$. Our assumed TMR of $110 \%$ corresponds to   $P \sim 0.6$, giving a $ \lvert \delta V_i  \rvert \sim 75  \ mV $.

Note that in this analysis the power drawn from $V_{IN}$ is not considered which is expected to be very different for a low input impedance design (BSN-A) compared to a high input impedance design (BSN-B) and will depend on the driving mechanism and circuitry. Overall, both designs suffer from significant steady-state power losses and would need to be turned off when not in use. This can be done straightforwardly for BSN-B using a large negative input voltage $V_{IN}$. The key point to note is that the energy dissipated during the evaluation of the BSN function is $\sim 20 \ \mu W \times 50 \ ps=$1 fJ which is orders of magnitude smaller than CMOS implementations of the same function \cite{ardakani2017vlsi, yuan2017vlsi} as noted earlier from system level simulations in \cite{zand2018composable}. The device level analysis presented here elucidates the role of proper magnet design for achieving the subnanosecond response times that is crucial for fast and low energy operation. The analysis also suggests low barrier IMA magnet as a more suitable candidate for BSN type applications due to its fast fluctuation dynamics, while modern non-volatile MRAM technology is largely based on PMA magnets \cite{bhatti2017spintronics}. 

\clearpage
\vspace{-10pt}
\footnotesize
\section*{Acknowledgment}
\vspace{-2.0pt}
This work was supported in part by the Center for Probabilistic Spin Logic for Low-Energy Boolean and Non-Boolean Computing  (CAPSL), one of the Nanoelectronic Computing Research (nCORE) Centers as task 2759.005, a Semiconductor Research Corporation (SRC) program sponsored by the NSF through ECCS 1739635.
\vspace{-10pt}
\bibliographystyle{IEEEtran}
\balance\bibliography{LBM_BSN_Hassan_Final}
 \end{document}